\documentclass{jfm}
\usepackage{graphicx}
\usepackage{natbib}
\usepackage{multirow}
\usepackage{epstopdf}
\usepackage{amssymb,amsmath}

\ifCUPmtlplainloaded \else
  \IfFileExists{amsbsy.sty}
{\typeout{^^JFound the 'amsbsy' package on the system, using
it.^^J}%
     \usepackage{amsbsy}}
    {\providecommand\boldsymbol[1]{\mbox{\boldmath $##1$}}}
\fi

\providecommand\bnabla{\boldsymbol{\nabla}}

\newsavebox{\astrutbox}
\sbox{\astrutbox}{\rule[-5pt]{0pt}{20pt}}

\title{Electroosmotic flow through a nanopore}

\author[M. Mao, J. D. Sherwood and S. Ghosal]%
{M. Mao$^1$
J. D.  Sherwood$^3$
 and
S. Ghosal$^{1,2}$%
\thanks{Email address for correspondence:
s-ghosal@u.northwestern.edu},\ns
\break
}

\affiliation{ $^1$Department of Mechanical Engineering and $^2$Engineering Sciences and Applied Mathematics, Northwestern University, 2145 Sheridan Rd, Evanston, IL 60208, USA\\[\affilskip]
$^3$Department of Applied Mathematics and Theoretical Physics, University of Cambridge, Wilberforce Road, Cambridge, CB3 0WA, UK}

\pubyear{}
\volume{}
\pagerange{}
\date{?; revised ?; accepted ?. - To be entered by editorial office}
\begin{document}
\maketitle

\begin{abstract}
Electroosmotic pumping of fluid through a nanopore that traverses an insulating membrane is considered. The density of surface charge on the membrane is assumed uniform, and sufficiently low for the Poisson-Boltzmann equation to be linearized. The reciprocal theorem gives the flow rate generated by an applied weak electric field, expressed as an integral over the fluid volume. For a circular hole in a membrane of zero thickness, an analytical result is possible up to quadrature. For a membrane of arbitrary thickness, the full 
Poisson--Nernst--Planck--Stokes system of equations is solved numerically using a finite volume method. The numerical solution agrees with the standard analytical result for electro-osmotic flux through a long cylindrical pore when the membrane thickness is large compared to the hole diameter. When the membrane thickness is small, the flow rate agrees with that calculated using the reciprocal theorem.
\end{abstract}

\begin{keywords}
\end{keywords}

\section{Introduction}
\label{sec:intro}
A nanopore is simply a hole of small size in an impermeable membrane separating two regions containing an electrolytic buffer. A size range of 1--100 nm is fairly typical. Living cells and intracellular organelles are usually bounded by lipid membranes containing nanopores
constructed of membrane-bound proteins. The transport of small molecules and polymers across such nanopores is a very common feature in living cells and is essential to their normal
function \citep{Alberts,Pfanner_90,Matouschek,Martin,Kunkele}.
Synthetic nanopores \citep{li_nat_mat03,storm_physRevE05,storm_nanolett05,dekker_nano_lett06, bayley_nanotechnology_2010,Garaj2010,Schneider2010} have been the focus of much interest in recent years following the demonstration of their use as effective single molecule sensors \citep{Kasianowicz1996}.

The main distinguishing feature of nanopore systems responsible for many of the novel effects is that their geometric dimensions are small enough so
that electrokinetic effects are important. Such effects have been invoked to explain a range of observations relating to experiments involving free as well as hindered translocation of DNA across synthetic nanopores \citep{Keyser2006,ghosal2006electrophoresis,ghosal2007effect,Ghosal2007,VanDorp2009,ghosal2013Nanoletter}. Nanopores also exhibit other unusual properties, some of which
could be potentially exploited to build novel microfluidic
devices. For example, conical nanopores fabricated on plastic films using
the ion track-etching technique \citep{siwy2006} have been shown to exhibit ion current rectification similar to that of semiconductor diodes
\citep{Siwy2004,Vlassiouk2007,Vlassiouk2008,Vlassiouk2008a}. A similar effect has been reported recently for electroosmotic flow out of nanocapillaries \citep{ghosal2013Nanoletter}. 
A nonlinear electrokinetic effect known as Induced Charge
Electroosmosis (ICEO) \citep{murtsovkin96, Squires2004}
produces vortices at the edges of nanopores resembling recirculation vortices in separated flows even though the Reynolds number in such applications is essentially zero.
Mixing due to these flow structures \citep{Yossifon2008,chang_understanding_2009-1,chang_nanoscale_2012} 
together with electroconvective instabilities \citep{ZALTZMAN2007} 
are thought to be responsible for the ``overlimiting'' behaviour of the 
current--voltage characteristics of perm selective pores and membranes
described by \citet{Rubinstein1979}.
Similar vortical structures may be generated in cylindrical channels that undergo a sudden 
constriction when the relevant length scales are on the order of the Debye length~\citep{park_eddies_2006}.

If an electric field $E$ is applied across an uncharged membrane, Induced Charge Electroosmosis leads to velocities $O(E^2)$ with no net flow through the membrane unless symmetry is somehow broken. Molecular dynamics simulations of flow through nanopores in uncharged membranes, such as graphene sheets \citep{Guohui2012}, show that differences in mobility between cations and anions can result in asymmetric Debye layers and consequent net flow through the membrane. Symmetry is also broken if the membrane is charged, so that the intrinsic field due to the membrane competes with the externally applied field in determining the distribution of ions in the Debye layer \citep{Mao2013}.

If a voltage is applied across a charged membrane containing a long narrow pore an electroosmotic flow is generated by the electric field acting on the charge cloud of counter-ions in the fluid adjacent to the fixed charges at the solid/fluid interface. The strength of this flow is proportional to the applied electric field, and thus, inversely proportional to the membrane thickness if the membrane is sufficiently thick. However, the flow does not increase indefinitely as the membrane is progressively thinned. If the membrane is much thinner than the diameter of the hole, the flow is driven mainly by electroosmosis at the membrane surface exterior to the pore, rather than by electric forces within the pore itself. 

Here we present results for the flow rate through a pore in a charged membrane in the limit of a weak applied field. In section \ref{sec:zeroThick} we use the reciprocal theorem to calculate the flow rate through a
circular hole in a charged membrane of zero thickness. The result is obtained in terms of an integral which in general has to be evaluated numerically but can be obtained analytically when the pore size is much smaller than the Debye length. In section \ref{sec:fullSim} we present computer simulations of the full problem based on numerical solutions of the Poisson-Nernst-Planck-Stokes system of equations using a finite volume method. Both thick and thin membranes are considered and compared with analytical results for membranes of zero thickness and of thickness large in comparison to the pore radius. Conclusions are provided in section~\ref{sec:conclusion}. 

\section{Flow through a hole in a membrane of zero thickness}
\label{sec:zeroThick}

\begin{figure}
\centering
\includegraphics[width=0.8\textwidth]{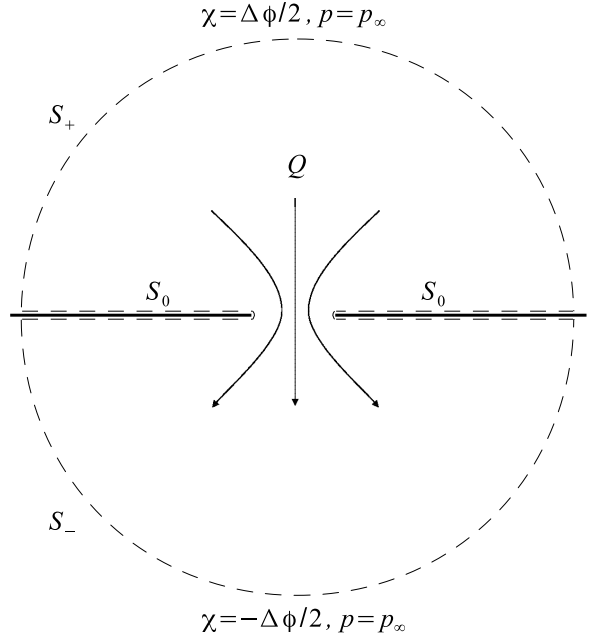}
\caption{Flow through a charged membrane under an applied potential difference $\Delta\phi$.}
\label{fig:sketch}
\end{figure}

We consider a hole (of arbitrary shape) in an infinite plane membrane immersed in an incompressible homogeneous electrolyte containing $N$ ionic species (Figure~\ref{fig:sketch}). The number density of the $i$th ionic species is $n^i$, with $n^i=n_\infty^i$ in the bulk electrolyte far from any charged surfaces. The electrolyte has viscosity $\mu$ and electrical permittivity $\epsilon$. A surface charge of fixed density $\sigma$ exists at the membrane/electrolyte interface, and in the electrolyte adjacent to the membrane there is charge cloud of counter ions, with thickness characterized by the Debye length
\begin{equation}
\kappa^{-1}=\left(
\frac{\epsilon kT}{\sum_{i=1}^N e^2 z_i^2n_\infty^i}\right)^{1/2},
\end{equation}
where $k$ is the Boltzmann constant, $T$ the absolute temperature, $e$ the proton charge, and $z_i$ the valence of the $i$th species of ion. We assume that the surface charge density $\sigma$ is sufficiently small so that the Poisson-Boltzmann equation describing the potential $\phi_0$ in the equilibrium charge cloud may be linearized. Far from any hole in the membrane, the zeta potential at the surface of the membrane is $\zeta=\sigma/(\epsilon\kappa)$,
with $\zeta\ll kT/e$. An electric potential difference $\Delta \phi$ is applied across the membrane with a resulting current $I$. The hole within the membrane has characteristic size $a$, and we make no assumption concerning $a\kappa$, the ratio of the hole size to the Debye length $\kappa^{-1}$. When $a\kappa\ll 1$, the charge clouds from opposite sides
of the perimeter of the hole overlap, but this has little effect upon the electrical conductivity of the hole when (as here) the surface charge density $\sigma$ and the resulting perturbations to the ionic number densities $n^i=n_\infty^i\exp(-ez_i\phi/kT)$ are small. Similarly, any ion exclusion effects of the overlapping charge cloud are negligible. We also note that the ion exclusion properties of a thin membrane are in general smaller than those of a long cylindrical pore. When $a\kappa\ll1$, the potential in the unperturbed electrical double layer within the pore, on the plane of the membrane, is $\zeta\approx\sigma/(\epsilon\kappa)$. Inside a uniform cylindrical pore with surface charge density $\sigma$, the potential when $a\kappa\ll 1$ is 
$\zeta\approx 2\sigma/(\epsilon a\kappa^2)$, and the condition $e\zeta/kT\ll 1$ required for the perturbation of the ionic number densities to be small within the pore implies a smaller charge density $\sigma$ for the pore than for the membrane. Only when $a\kappa=O(1)$ will the long cylindrical pore and the hole in a membrane exhibit similar ion exclusion properties.

When $a\kappa\gg 1$ the charge cloud is thin compared to the lateral dimension of the hole, but it remains thick compared to the membrane of zero thickness $h=0$ considered in \S~\ref{subsec:roundhole}. Thus, we are unable to appeal to Smoluchowski's analysis for thin charge clouds in \S~\ref{subsec:roundhole}. Ion exclusion effects are negligible in this limit.

In \S~\ref{subsec:charge_perturbation} we discuss how the charge cloud is deformed by the applied electric field and by fluid motion. We then (\S~\ref{subsec:formalism}) describe a theoretical framework for calculating the flow rate $Q$ through the hole, exploiting the reciprocal theorem: the analysis is similar to that of \citet{Sherwood1995}.  In \S~\ref{subsec:roundhole} we consider the special case of a circular hole in a thin membrane for which the integral for the flow rate $Q$ can be computed numerically.

\subsection{\label{subsec:charge_perturbation}The perturbed charge cloud}

When the electrical potential difference $\Delta\phi$ is applied across the membrane, the charge cloud adjacent to the surface of the membrane is perturbed, both by the direct electrical field $\nabla\phi$
acting on the ions, and by motion of the fluid. Ions are convected with the fluid velocity $\mathbf{u}$, and move relative to the fluid under the influence of electric fields and thermal diffusion.
The conservation equation for the number density $n^i$ of the $i$th ionic species, in steady state, is therefore
\begin{equation}
\nabla\cdot\left\lbrack n^i\mathbf{u} -\omega^i(kT\bnabla
n^i + ez_in^i\bnabla\phi) \right\rbrack=0 ,
\end{equation}
where $\omega^i$ is the mobility of the $i$th species of ion.

We follow \citet{saville1977} and nondimensionalise potentials by $kT/e$, lengths by the typical hole dimension $a$, velocities by $\epsilon(kT/e)^2/\mu a$ and mobilities by a characteristic mobility
value $\omega^0$. We assume that the potential $\Delta\phi$ which characterises the applied field is small compared to the equilibrium zeta potential $\zeta$, so that $\beta=e\Delta\phi/kT\ll e\zeta/kT$,
where $e\zeta/kT$ has already been assumed small in order that we may describe the charge cloud by means of the linearized Poisson-Boltzmann equation. We use the dimensionless field strength $\beta$ as the basis for a perturbation expansion:
\begin{subeqnarray}
\hat{\mathbf{u}} &=& \beta\hat{\mathbf{u}}_1 + \cdots\\
\hat\phi &=& \hat\phi_0 + \beta\hat\phi_1 +\cdots\\
n^i &=& n_0^i + \beta n_1^i + \cdots,
\end{subeqnarray}
where the subscript 0 refers to the equilibrium cloud, and the caret $\hat{ }$ denotes a nondimensional quantity.  The steady-state ion conservation equation, correct to $O(\beta)$, becomes
\begin{equation}
\text{Pe} ~\hat{\mathbf{u}}_1\cdot\bnabla n_0^i = \hat\omega^i\bnabla\cdot\left\lbrack
 z_in_0^i\bnabla\hat\phi_1 + z_in_1^i\bnabla\hat\phi_0 +\bnabla
n_1^i \right\rbrack
\label{convdiff}
\end{equation}
where the Peclet number $\text{Pe}=\epsilon kT/e^2\mu\omega^0$  characterizes the ratio of ionic convection to diffusion. Since the nondimensional equilibrium potential $\hat\phi_0$ has been assumed to be small, (\ref{convdiff}) reduces to
\begin{equation}
0 = z_in_\infty^i\nabla^2\hat\phi_1 + \nabla^2n_1^i . \label{eqn1}
\end{equation}
The boundary conditions at infinity are
\begin{subeqnarray}
n_1^i&\to&0,\\
\beta\phi_1&\sim&\pm\Delta\phi/2\hbox{\hskip 10pt in $z\gtrless 0$.}
\label{bc_infinity}
\end{subeqnarray}
We assume that no ions enter or leave the surface of the membrane. Hence
\begin{equation}
\mathbf{n}\cdot(kT\bnabla n^i+ez_in^i\bnabla\phi)=0,
\end{equation}
where $\mathbf{n}$ is the normal to the membrane. At $O(\beta)$, and assuming $\hat\phi_0\ll1$, this zero flux boundary condition becomes
\begin{equation}
\mathbf{n}.\bnabla \left(n_1^i+z_in_\infty^i\hat\phi_1\right)=0 . \label{bc1}
\end{equation}

Multiplying equation (\ref{eqn1}) by $e^2z_i$ and summing over $i$, we obtain, at $O(\beta)$,
\begin{equation}
\nabla^2\hat\chi_1=0,
\label{chieqn1}
\end{equation}
where
\begin{equation}
\hat\chi_1 = \hat\phi_1 + \hat\rho_1(a\kappa)^{-2},
\label{chieqn2}
\end{equation}
and
\begin{equation}
\hat\rho_1=\frac{ea^2\rho_1}{\epsilon kT}=\frac{ea^2}{\epsilon kT}\sum_{i=1}^N
ez_in_1^i
\end{equation}
is the (non-dimensional) perturbation to the charge density
$\rho=\sum_iez_in^i$.

The boundary conditions for (\ref{chieqn1}) are similarly obtained from
 (\ref{bc_infinity}) and (\ref{bc1}):
\begin{subeqnarray}
\hat\chi_1&\sim&\pm 1/2\quad
\hbox{as $\mathbf{r}\rightarrow\infty$ in $z\gtrless 0$,}
\\
\mathbf{n}\cdot\bnabla\hat\chi_1&=&0\hskip 35pt\hbox{on the membrane,}
\label{chieqn3}
\slabel{chie_normal_bc}
\end{subeqnarray}
and we note that the boundary condition (\ref{chie_normal_bc}) represents zero flux of ions into the membrane, rather than a zero normal electric field. The potential
\begin{equation}
\chi=\phi_1+\rho_1/\epsilon\kappa^2
\end{equation}
is thus obtained by solving the Laplace equation for the potential created by applying the potential difference $\Delta\phi$ across the insulating membrane containing the pore. The potential $\chi$ is related to the linearized change
$\mu_1^i=ez_i\phi_1+kTn_1^i/n_\infty^i$ in the electrochemical potential of the $i$th species of ion, with $\chi=\sum_iez_in_\infty^i\mu_1^i/(\epsilon kT\kappa^2)$. As discussed by \citet{saville1977},
when $\hat\phi_0\ll 1$ the perturbation $\rho_1$ to the charge density
is negligibly small, so that $\phi_1=\chi$, and our analysis is equivalent to that of~\citet{Henry_1931}.

\subsection{A formalism for calculating the flow rate using the
reciprocal theorem}
\label{subsec:formalism}

The Stokes equations governing fluid motion are modified by the presence of an electric force $-\rho\nabla\phi$ acting on the fluid. Expanding in powers of $\beta$, we find
\begin{eqnarray}
\rho\bnabla\phi&=&(\rho_0+\beta\rho_1)\bnabla(\phi_0+\beta\phi_1)+O(\beta^2)
\nonumber\\
&=&-\bnabla(\hbox{$\frac{1}{2}$}\epsilon\kappa^2\phi_0^2-\beta\rho_1\phi_0)
+\beta\rho_0\bnabla(\phi_1+\rho_1/\epsilon\kappa^2)+O(\beta^2),
\label{rhophi}
\end{eqnarray}
where we have used the relation $\rho_0=-\epsilon\kappa^2\phi_0$ between the
equilibrium charge density $\rho_0$ and the equilibrium potential $\phi_0$ given by the linearized Poisson-Boltzmann equation. Hence the Stokes equations become
\begin{equation}
\mu\nabla^2\mathbf{u} -\bnabla p - \rho_0\bnabla\chi = 0,
\label{stokes}
\end{equation}
where the term $\bnabla(\frac{1}{2}\epsilon\kappa^2\phi_0^2-\beta\rho_1\phi_0)$ in
(\ref{rhophi}) has been incorporated into the pressure $p$.

Note that the fluid motion caused by direct electrical forces acting on the fluid  create additional deformation of the charge cloud, but, as seen from equation (\ref{convdiff}), this deformation is $O({\hat u}_1 \text{Pe}\; \hat\phi_0)$, and may be neglected when
$\hat\phi_0$ is small. 

We now determine the $O(\beta)$ total volumetric flow rate $Q$ through the pore, created by the electrical force acting on the charge cloud within the fluid. Consider two Stokes flows $\mathbf{u}$ and $\bar{\mathbf{u}}$ in a volume $V$ with boundary conditions given on the bounding surface $S$
with outward normal $\mathbf{n}$. A body force $\mathbf{F}$ acts on the fluid, in which the pressure is $p$, the viscosity is $\mu$, the strain rate tensor is $e_{ij}=(\partial_iu_j+\partial_ju_i)/2$, and the stress tensor is $\tau_{ij}= - p \delta_{ij} + 2\mu e_{ij}$. Barred variables represent the corresponding quantities for the second flow $\bar{\mathbf{u}}$. The reciprocal theorem \citep{Happel&Brenner} gives the identity:
\begin{equation}
\int_V u_i \bar{F}_i dV + \int_S u_i \bar{\tau}_{ij} n_j dS = \int_V
\bar{u}_i F_i  dV + \int_S \bar{u}_i \tau_{ij} n_j dS.
\label{eq:recTheo}
\end{equation}

We suppose that flow 1 ($\mathbf{u}$)  is the flow of interest, namely, the electrokinetic flow through a hole in a charged membrane. The body force $\mathbf{F}$, by (\ref{stokes}), is
\begin{equation} 
\mathbf{F} = -\rho_0\bnabla\chi,
\label{eq:forceF}
\end{equation} 
where, by (\ref{chieqn1}), $\nabla^2 \chi = 0$ with boundary conditions (\ref{chieqn3})
$\hat{\mathbf{n}} \cdot \bnabla \chi = 0$ on the membrane surface $S_0$ and  $\chi \sim \pm \Delta \phi / 2$ on surfaces $S_{\pm}$ far from the pore (Figure~\ref{fig:sketch}).

We take flow 2 ($\bar{\mathbf{u}}$) to be that due to a pressure difference $\Delta p$ imposed across a pore in an uncharged membrane; thus, the body force $\bar{\mathbf{F}}=0$. Since the Stokes flow equations are linear, we may write 
\begin{equation} 
\bar{\mathbf{u}} = \Delta p\, \mathbf{G},
\label{eq:defineG}
\end{equation} 
where $\mathbf{G}$ is a function that depends solely on the pore geometry. 

We now substitute the two flows into the reciprocal relation (\ref{eq:recTheo}). The bounding surface $S$ is as shown in Figure~\ref{fig:sketch}; it consists of two hemispheres $S_{+}$ and $S_{-}$ of very large radius $R$, together with the membrane surface, $S_{0}$.
On $S_{0}$, the velocities $u_{i}, \bar{u}_{i}=0$ whereas on 
$S_{\pm}$ we have $\tau_{ij} \sim - p_{\pm} \delta_{ij} + O(R^{-3})$ and 
$\bar{\tau}_{ij} \sim - \bar{p}_{\pm} \delta_{ij} + O(R^{-3})$ where $p_{\pm}$ are the pressures 
at great distance $R$ from the pore on either side of the membrane. 
For flow 1, $p_{+}=p_{-} = p_{\infty}$ and for flow 2, $\bar{p}_{+} - \bar{p}_{-} = \Delta p$. Substituting
(\ref{eq:forceF}) and (\ref{eq:defineG}) into (\ref{eq:recTheo}), and cancelling the pressure difference $\Delta p$ from both sides of the equation, we obtain
\begin{equation}
Q = \int_{S_-}\mathbf{u}\cdot\mathbf{n}\; dS=
 - \int_{S_+}\mathbf{u}\cdot\mathbf{n}\; dS =
- \int_V \rho_0 \mathbf{G} \cdot \bnabla \chi \; dV,
\label{eq:Q1}
\end{equation}
where $Q$ is the volumetric flux of flow 1 from the side $S_{+}$ to the side $S_{-}$ of the 
membrane.

\subsection{Flow rate from a round hole in elliptic cylindrical co-ordinates}\label{subsec:roundhole}
We now consider a circular pore of radius $a$. We adopt cylindrical co-ordinates  $(r,z)$, with origin at the centre of the pore and $z$ along the axis of symmetry, together with oblate spherical coordinates $(\xi,\eta)$ where $\infty > \xi > - \infty$, $\pi/2 > \eta \geq 0$ such that
\begin{equation}
z=a\sinh\xi\cos\eta\quad,\quad r=a\cosh\xi\sin\eta.
\end{equation}
The scale factors are
\begin{equation}
h_\xi=h_\eta=a(\cosh^2\xi-\sin^2\eta)^{1/2}.
\end{equation}
The imposed electric field is given by  \citet[p. 1292]{M&F}, with potential
\begin{equation}
\chi = \frac{\Delta \phi}{2} \left[ 1- \frac{2}{\pi}
\tan^{-1}\left( \frac{1}{\sinh \xi} \right) \right].
\label{eq:chi}
\end{equation}
\citet[p. 153]{Happel&Brenner}
give the stream function $\psi=-a^3\Delta p(1-\cos^2\eta)/(6\pi\mu)$ for
flow 2. Comparing the resulting velocity, $\bar{\mathbf{u}}$, with
(\ref{eq:defineG}),
\begin{equation}
G_\xi =  - \frac{a \cos^2\eta}{2\pi\mu
\cosh\xi(\cosh^2\xi-\sin^2\eta)^{1/2}}\quad,\quad
G_\eta  =  0.
\label{eq:u2}
\end{equation}
Substituting (\ref{eq:chi}) and (\ref{eq:u2}) into (\ref{eq:Q1}) yields the electroosmotic flow rate
\begin{equation}
Q   =  \frac{2a^3\Delta\phi}{\pi\mu}
\int_0^{\frac{\pi}{2}} d\eta \int_0^\infty \rho_0
\frac{\cos^2\eta\sin\eta}{\cosh\xi} d\xi.
\label{eq:Q2}
\end{equation}
The equilibrium charge density $\rho_0$ in the linearized, Debye H\"{u}ckel limit may be obtained by excising the solution for a uniformly charged disk  \citep{Sherwood1995} from that for a charged infinite plate. 
Hence
\begin{eqnarray}
\rho_0 & = & \sigma  \kappa^2 a \left[  \int_0^\infty
\frac{J_1(as)J_0(rs)}{(\kappa^2+s^2)^{1/2}}
e^{-(\kappa^2+s^2)^{1/2} z} ds - \frac{e^{-\kappa z}}{\kappa a} \right].
\label{eq:rho0}
\end{eqnarray}

The integral in (\ref{eq:Q2}) cannot be evaluated in closed form when $\rho_0$ is given by (\ref{eq:rho0}). However, in the long Debye length limit  $\kappa a \ll 1$, the rate of decay of $\mathbf{G}$ and $\bnabla \chi$ is such that the major contribution to the integral (\ref{eq:Q2}) comes from a region (near the hole) of volume $O(a^3)$, within which $\rho_0 \approx -\sigma\kappa$. Thus,
\begin{equation}
Q \sim - \frac{2a^3\Delta\phi}{\pi\mu} (\sigma \kappa)
\int_0^{\pi/2} \cos^2\eta \sin\eta \, d\eta \int_0^\infty
\frac{d\xi}{\cosh\xi} = - \frac{a^3\kappa\sigma\Delta\phi}{3\mu} = \kappa a Q_{0},
\label{eq:QaKappaSmall}
\end{equation}
where $Q_{0}= - a^2 \sigma\Delta\phi/ (3\mu)$ is a convenient characteristic flow rate.

In other cases, the integral must be evaluated numerically. It is convenient to introduce new variables 
$x=\kappa a$, $t=sa$, $\bar{r} = r/a$, $\bar{z} = z/a$ and $q=\cos\eta$ in terms of which (\ref{eq:Q2}) becomes 
\begin{equation}
Q = \frac{2a^2 \sigma \Delta\phi}{\mu\pi} \left[ - x
I_2 + x^2 \int_0^{\pi/2} d\eta \int_0^\infty I_1
\frac{\cos^2\eta\sin\eta}{\cosh\xi} d\xi \right],
\label{eq:Q3}
\end{equation}
with $I_1$ and $I_2$ defined as 
\begin{eqnarray} 
I_1 &=& \int_0^\infty
\frac{J_1(t)J_0(\bar{r}t)}{\sqrt{x^2+t^2}}
\exp[ - \bar{z} \sqrt{x^2+t^2}   ] dt, 
\label{eq:I1_define}
\\ 
I_2 &=&  \int_0^1 q^2 \left[
ci(x q)\sin(x q )- si(x q )\cos(x q) \right] d q,
\label{eq:expTermInt}
\end{eqnarray}
where $si(\alpha), ci(\alpha)$ are the sine and cosine integrals 
\begin{equation}
si(\alpha) = - \int_\alpha^\infty \frac{\sin t}{t}\, dt\quad , \quad
ci(\alpha) = -\int_\alpha^\infty \frac{\cos t}{t}\,dt.
\end{equation}
\begin{figure}
\centering
\includegraphics[width = 0.8\textwidth]{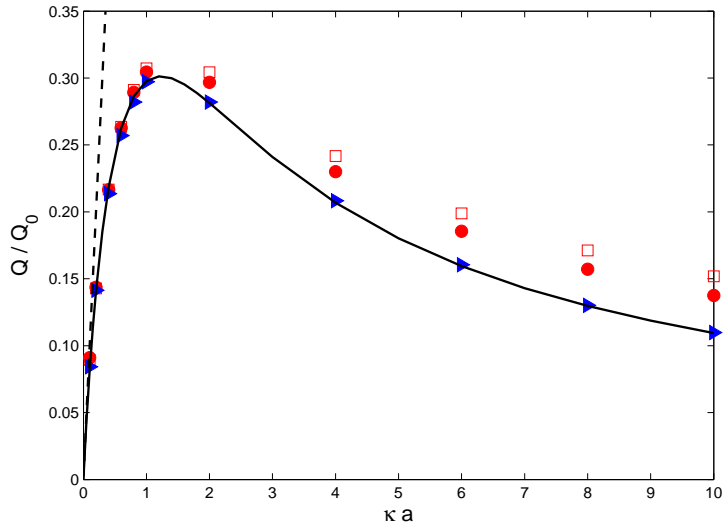}
\caption{The normalized flow rate $Q/Q_0$ through a circular pore of radius $a$ 
in a membrane of thickness $h=0$ as a function of $\kappa a$ determined from (\ref{eq:Q3})  (solid line), with asymptote $Q/Q_{0} \sim \kappa a$ (\ref{eq:QaKappaSmall}) (dashed line). The symbols are from the full finite volume simulations of \S~\ref{sec:fullSim} with $h/a=$ 0 (triangle), 0.06 (circle) 
and 0.1 (square).}
\label{fig:QaKappa}
\end{figure}

The integrals $I_1$ (\ref{eq:I1_define}) and $I_2$ (\ref{eq:expTermInt}) were evaluated using the Matlab routine \textbf{quadgk} \citep{matlab}. $I_1$ represents the potential due to a thin charged disk,
and decays exponentially at large distances, as does $1/\cosh(\xi)$. The $\xi$ integration in
(\ref{eq:Q3}) could therefore be truncated at a large value, taken to be $\xi=8$
(corresponding to a distance $\approx 1490a$ from the pore). To evaluate the second term in brackets in (\ref{eq:Q3}), the $(\xi,\eta)$ space was divided into sub--regions, with typically 500 intervals for $\xi$ and 200 for $\eta$. Smaller intervals were used near the pore ($\xi\ll 1$) and near
the membrane ($\pi/2-\eta\ll 1$). For every pair $(\xi,\eta)$, the integral $I_1(\xi,\eta)$ was numerically
evaluated and $Q$ was obtained via trapezoidal summation within Matlab. Results are shown in Figure~\ref{fig:QaKappa} with the asymptotic regime $\kappa a \ll 1$, described by (\ref{eq:QaKappaSmall}), 
depicted by the dashed straight line of unit slope.

\section{Flow through a hole in a membrane of finite thickness}
\label{sec:fullSim}
In order to examine the validity of (\ref{eq:Q3}), the full Poisson-Nernst-Planck-Stokes system of equations was solved numerically using a finite volume method based on the open source CFD library OpenFOAM \citep{OPENFOAM}. Our model has been described by \citet{Mao2013} and details of the implementation for the current problem are presented in the Appendix. The cases studied include membranes of thickness $h=0$ and $h>0$ in the regime of weak applied fields and low membrane charge. These conditions, stated in \S~\ref{subsec:formalism}, may be restated more conveniently as $ \hat{\phi}^{-1} \gg \kappa a \gg \hat{\sigma}$ where $\hat{\sigma} = a e |\sigma| / (\epsilon kT)$ and $\hat{\phi} = |\epsilon\Delta \phi / (\sigma a)|$ are dimensionless parameters characterizing the degree of membrane charge and the strength of the applied field respectively. In the simulations presented here, the values of these parameters were $\hat{\sigma} = 0.273$ and $\hat{\phi}=0.071$ so that (\ref{eq:Q3}) can be expected to be a reasonable approximation in the range $14 \gg \kappa a \gg 0.3$. 

The computed flow rate $Q$ normalized by $Q_{0} =  - a^2 \sigma\Delta\phi/(3\mu)$ is shown 
by the symbols in Figure~\ref{fig:QaKappa}. Good agreement with (\ref{eq:Q3}) is obtained, but  
thicker membranes result in somewhat increased flow rates. The discrepancy increases at 
shorter Debye lengths.  

The analysis of \S~\ref{sec:zeroThick} assumed that effects due to induced charge electroosmosis (ICEO) are negligible. However, \citet{Thamida2002} have shown that ICEO generates vortices at sharp corners, and such
vortices will inevitably be generated in the membrane geometry considered here. In the vicinity of the edge of the nanopore, where $r/a=1+s$, the potential $\chi$ (\ref{eq:chi}) on the membrane surface
$z=0$ may be expanded as
\begin{equation}
\chi=\pm\frac{\Delta\phi}{\pi}\sqrt{2s},\quad \hbox{on $z=\pm0$.}
\end{equation}
If we assume that this potential is little modified when the membrane has a finite thickness $h>0$, the potential gradient within the solid membrane due to the external potential $\chi$ is
\begin{equation}
\frac{\partial\phi_s}{\partial z}=\frac{2\Delta\phi}{\pi}\frac{\sqrt{2s}}{h},
\end{equation}
and if $\epsilon_s\ll\epsilon$ the induced potential gradient normal to the surface within the liquid is 
$(\epsilon_s/\epsilon)\partial\phi_s/\partial z$. This corresponds to an induced surface charge (with accompanying charge cloud of couter-ions)
\begin{equation}
\sigma_i=\mp\frac{2\epsilon_s\Delta\phi}{\pi}\frac{\sqrt{2s}}{h}
,\quad \hbox{on $z=\pm0$.}
\label{induced_charge}
\end{equation}
The analysis breaks down when $s \lesssim h/a$ where the detailed geometry near the edge of the pore becomes important. If the membrane has rounded edges, the curvature $\sim h^{-1}$ and the induced charge is at most $\sigma_{i} \sim \epsilon_s\Delta\phi/(ah)^{1/2}$. This may be neglected as long as it is small compared to $\sigma$, or equivalently, if $(\epsilon_s/\epsilon) \sqrt{(a/h)} \; \hat{\phi} \ll 1$.
For common membrane materials (e.g. lipids and silica) $\epsilon_s / \epsilon \sim 0.1$, thus, 
as long as the applied field remains weak, ICEO effects are restricted to the neighbourhood of
sharp corners.

The results of full numerical solutions of the Poisson-Nernst-Planck-Stokes equations reported in Figure~\ref{fig:QaKappa} were computed with membrane permittivity $\epsilon_s=0$.
However, it is shown in the Appendix that numerical solutions for appropriate non-zero solid permittivities $\epsilon_s >0$ predict flow rates $Q$ that differ little from those with $\epsilon_s=0$
in the parameter regime under consideration. The contribution of such effects to the net fluid flux, $Q$,
is at best very weak. This is perhaps not surprising, since even when (as here) there are sharp corners, if the membrane is uncharged and the electrolyte is symmetric (with identical ionic mobilities),
symmetry dictates that ICEO cannot generate a net flow $Q$ through the pore. So although nonlinear
effects such as ICEO can generate a net fluid flow through a charged membrane, the applied field $\Delta\phi$ must be larger than the fields considered here. An ICEO contribution to the fluid flux is in principle possible, but only for a charged membrane at high applied fields, as, for example, in the numerical results presented by \citet{Mao2013}.

The reciprocal theorem used in \S~\ref{subsec:formalism} enabled
us to determine the volumetric flow rate $Q$ through the pore without a full computation of the velocity field. This has the advantage of leading quickly to a value for $Q$. However, this approach hides other interesting features of the flow, such as eddies, whether generated by ICEO \citep{Thamida2002} or by the pore throat restricting the flow \citep{park_eddies_2006}. These features are only revealed by full numerical computations, such as those discussed in the Appendix.

\subsection{The limits of thick and thin membranes} 
Since the fluid flux through the pore is generated by the applied potential, $\Delta \phi$,
we can define an ``electroosmotic conductance'' $H = Q/\Delta\phi$ in analogy to the electric conductance. If the membrane thickness  $h \gg a$, we have a long cylindrical pore 
with surface charge density $\sigma$ at the wall. 
The electroosmotic flow velocity is then \citep{levine1975}
\begin{equation}
u = \frac{\epsilon E_0}{\mu} [\phi_0 - \zeta]
\label{eq:axial_velocity_u}
\end{equation} 
where $\phi_0=-\epsilon\kappa^2\rho_0$ is the equilibrium potential in the double layer, $\zeta$ is the equilibrium potential at the wall, and $E_0 = \Delta\phi / h$. The equilibrium potential of a cylindrical pore in the Debye-H\"{u}ckel limit is
\begin{equation}
\phi_0 = \zeta \frac{I_0(\kappa r)}{I_0(\kappa a)}
= \frac{\sigma}{\epsilon\kappa} \frac{I_0(\kappa r)}{I_1(\kappa a)}.
\end{equation} 
Integrating the fluid velocity $u$ (\ref{eq:axial_velocity_u}) over the cross-section 
we obtain the volumetric flow rate $Q$ and hence the the electroosmotic conductance \citep{rice1965}
\begin{equation}
H_c = \frac{Q}{\Delta\phi} = \frac{2\pi\sigma a^3}{\mu h} \left[
\frac{1}{(\kappa a)^2} - \frac{1}{2(\kappa a)} \frac{I_0(\kappa
a)}{I_1(\kappa a)} \right].
\label{eq:Hc}
\end{equation}
On the other hand, when the thickness $h/a \ll  1$, we expect the system to be identical to a hole in a zero--thickness membrane, and the electroosmotic conductance may be obtained using (\ref{eq:Q3}):
\begin{equation}
H_p = \frac{Q}{\Delta\phi} =
\frac{2 \sigma a^2}{\mu\pi} \left[ - (\kappa a)
I_2 + (\kappa a)^2 \int_0^{\pi/2} d\eta \int_0^\infty I_1
\frac{\cos^2\eta\sin\eta}{\cosh\xi} d\xi \right].
\label{eq:Hp}
\end{equation} 

\subsection{The electroosmotic access resistance of a nanopore}
If a membrane of thickness $h$ containing a circular hole of radius $a$ separates two uniformly conducting regions, then the electrical resistance of the cylindrical hole increases,
proportional to $h$. This might suggest a vanishing resistance for an infinitely thin membrane. However, in reality, as $h\rightarrow 0$ the electrical resistance is dominated by entrance and exit effects,
and can be determined from the electrical potential (\ref{eq:chi}). This is called the ``access resistance'' of the pore and for a circular pore in an infinitely thin membrane it is described by a simple analytical formula \citep{Hall1975}.

An analogous situation applies to the problem of electroosmotic flow through a pore in a membrane. When the pore length $h$ is large compared to the pore radius $a$, the flow conductance $H \sim H_c \sim h^{-1}$ from (\ref{eq:Hc}) -- a consequence of the fact that the electric field in the pore, $E \sim \Delta \phi / h$. However, $H$ does not increase indefinitely as $h \rightarrow 0$ but instead approaches a finite value $H_p$ given by (\ref{eq:Hp}). The surface charge $2\pi ah\sigma$ within the cylindrical pore goes to zero as $h\rightarrow 0$. Electroosmotic motion is therefore determined by
flow in the fluid on either side of the membrane, as described in \S~\ref{sec:zeroThick}, and not by the cylindrical pore. Thus, in analogy to the corresponding electrical problem, $H_p^{-1}$ may be regarded as an  ``access resistance'' of the pore to electroosmotic flow.

\begin{figure}
\centering
\includegraphics[width=0.45\textwidth]{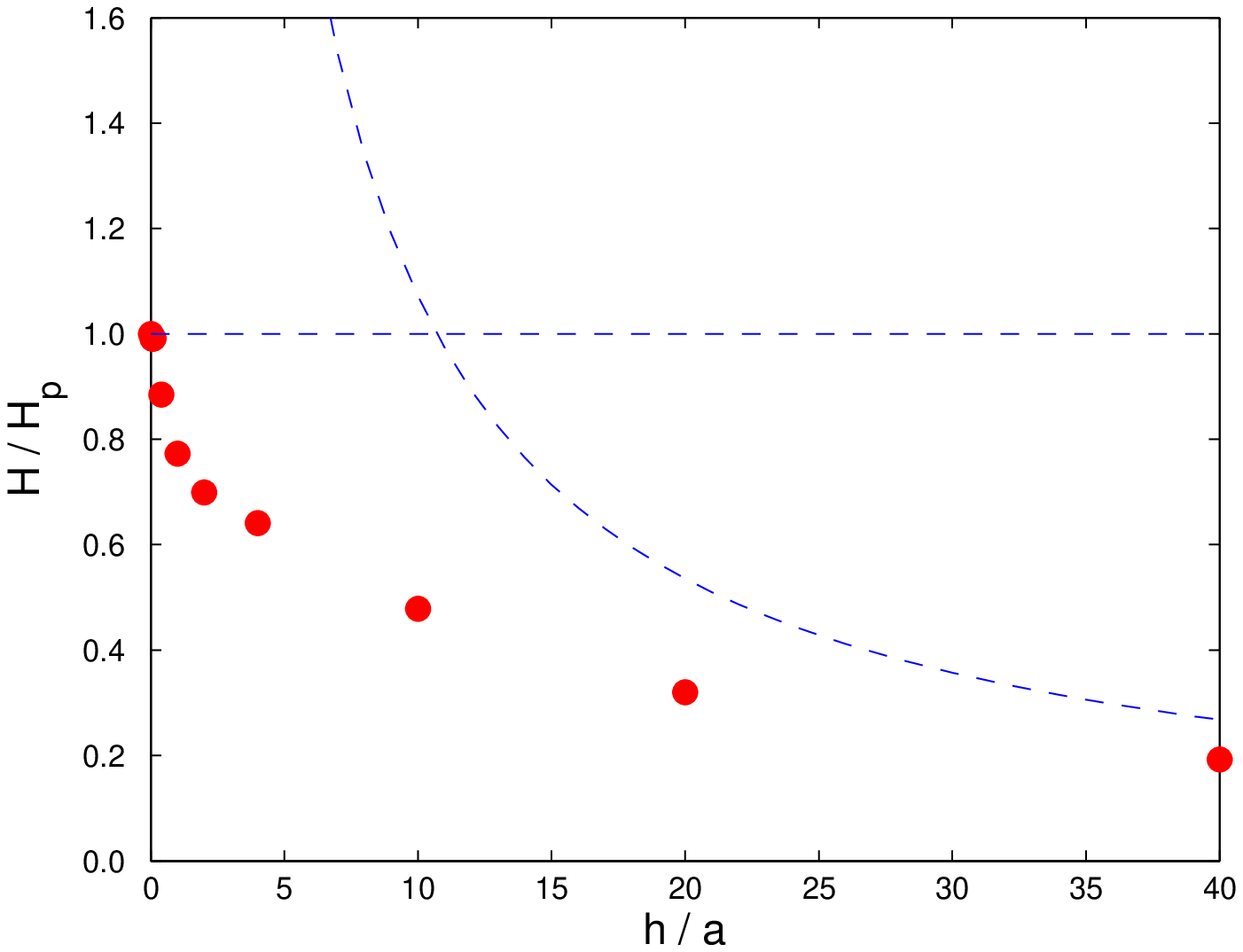}
\includegraphics[width=0.45\textwidth]{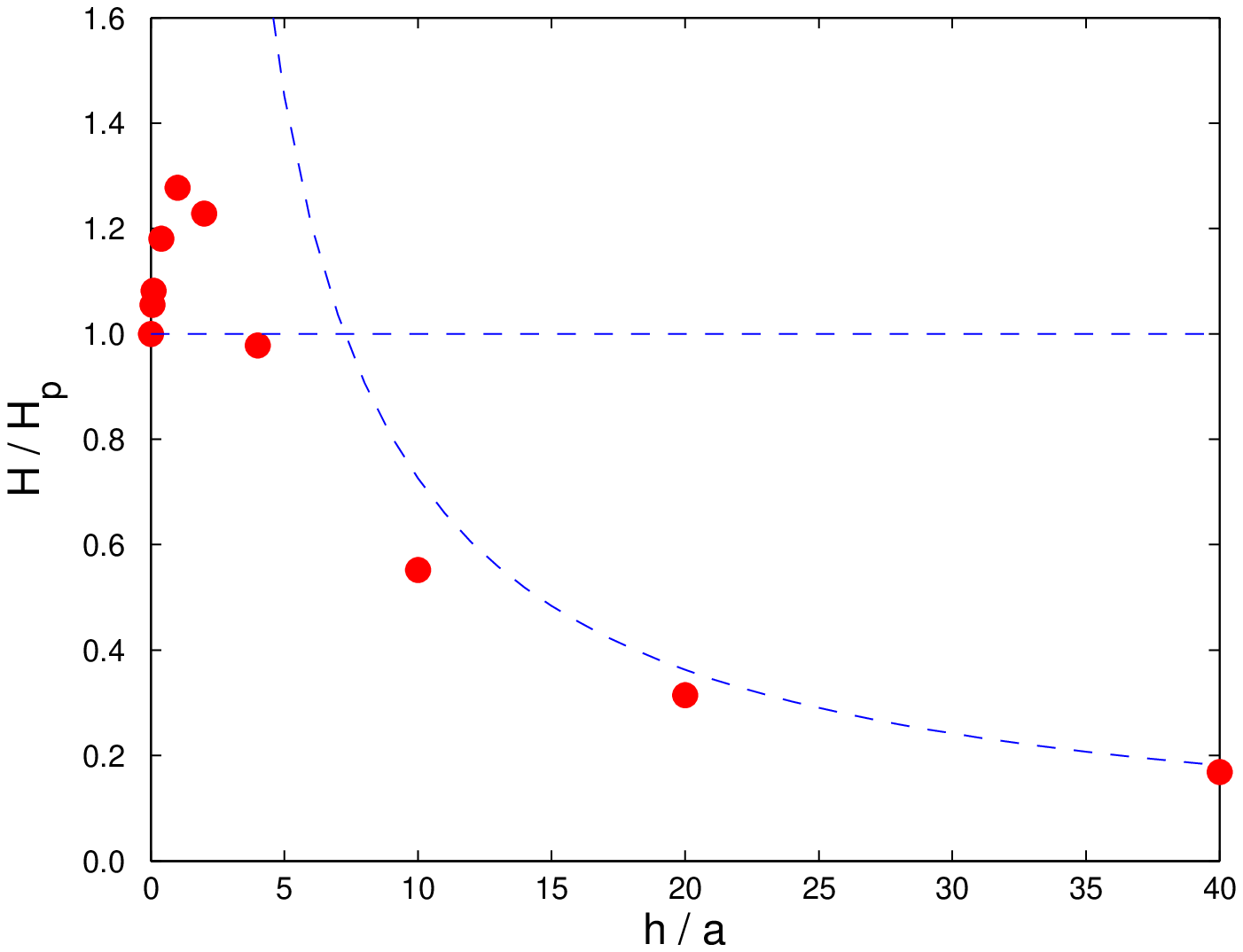}
\caption{The normalized ``electroosmotic conductance'' $H/H_p$ determined from the full 
numerical simulation (symbols) as a function of the normalized membrane thickness, $h/a$. Left panel corresponds to $\kappa a = 0.4$ and right panel $\kappa a = 2.0$.
The dashed lines correspond to the thin ($H=H_p$) and thick 
($H=H_c$) membrane limits obtained from (\ref{eq:Hc}) and (\ref{eq:Hp}).}
\label{fig:flowConduc}
\end{figure}

Figure~\ref{fig:flowConduc} shows the electroosmotic conductance $H$ obtained from 
the finite volume numerical computations as a function of  $h/a$. The computed value of $H$ is normalized by $H_p$ obtained from (\ref{eq:Hp}). It is seen that $H/H_p$ approaches unity as $h/a \rightarrow 0$ and approaches $H_c/H_p \sim h^{-1}$ for large $h/a$. The dashed line representing $H_c/H_p$ was obtained from (\ref{eq:Hc}) and (\ref{eq:Hp}). The results of the full computations indicate that though $H$ does exhibit the expected limiting behaviors, it does not vary monotonically with $h$ at short Debye lengths. The origin of the peak at intermediate values of $h/a$ will be investigated further in future work.

\section{Concluding Remarks}\label{sec:conclusion}

We have assumed that the surface charge density $\sigma$ is sufficiently low 
that the zeta potential is small, $\zeta\ll kT/e\approx 25\rm\ mV$ at $T=298\rm\ K$.
Thus, the Poisson-Boltzmann equation can be linearized.
Non-dimensional zeta potentials $e\zeta/kT$ in colloidal systems, though not always small, are typically at most 5, and it is found that theories based on small potentials usually give useful
qualitative insight into electrokinetic behaviour over this range of potentials (e.g. \citet{levine1975}).

We have also assumed that the applied potential difference $\Delta\phi\ll\zeta$. Potential differences applied in experiments are typically of the same order as typical $\zeta$-potentials
which in silica substrates vary in magnitude between $0$ and $100$ mV 
depending mainly on counter-ion concentration~\citep{zetareview_eph04a,zetareview_eph04b}. 
For example, \citet{Keyser2006} describe experiments in which $\Delta\phi$ was in the range 30--100~mV. Nanopores (radius $\sim$ 5--10 nm) in graphene sheets have recently been used in
DNA translocation experiments \citep{Garaj2010,Schneider2010,Merchant2010}.
The applied voltage $\Delta \phi \sim$ 0--200 mV in these experiments. The computations of \citet{Mao2013} predict that the electroosmotic flow rate through a pore in a membrane
varies non-linearly with $\Delta\phi$ only at voltages greater than 100 mV. Thus, we again expect the results presented here to give at least a qualitative understanding of electroosmotic flow in such experiments.

Electroosmotic flow through nanopores has been shown to control the translocation 
velocity of charged polymers in resistive pulse  
experiments~\citep{ghosal2006electrophoresis,ghosal2007effect}. When the free translocation of the polymer is hindered by tethering it to a colloid held in an optical trap, the tethering force has been shown to be determined by the electroosmotic flow within the pore~\citep{Ghosal2007,Keyser2006,laohakunakorn_dna_2013}.
Furthermore, it has been argued that the flow outside and in the vicinity of the nanopore 
controls the capture rate of polymers into the pore~\citep{wong_polymer_2007}, though the 
experimental evidence for this appears tentative at present.

In addition to the single molecule experiments mentioned above, our results should also be helpful 
in understanding the properties of nanoporous membranes that are used in batteries, water desalination and numerous other industrial applications. \citet{gadaleta_sub-additive_2014} have recently studied the electrical conductivity of a model membrane consisting of an array of nanopores. However, the applied voltage should also result in an electroosmotic flux, the calculation of which may be undertaken as a suitable generalization of the approach presented here. 
Since membrane bound organelles in cells contain nanopores that control the traffic of biological molecules across the membrane, our results may also be of interest in the biological 
context (see e.g.~\citet{gu_electroosmotic_2003}). 

\section*{Acknowledgement}
MM \& SG acknowledge support from the NIH through Grant 4R01HG004842-03. SG was hosted by the
Cavendish Laboratory, University of Cambridge, as visiting Professor with funds provided by the 
Leverhulme Trust. JDS thanks the Department of Applied Mathematics and Theoretical Physics, University of Cambridge, and the Institut de M\'ecanique des Fluides de Toulouse, for hospitality.

\appendix
\begin{figure}
\centering
\includegraphics[width=1.0\textwidth]{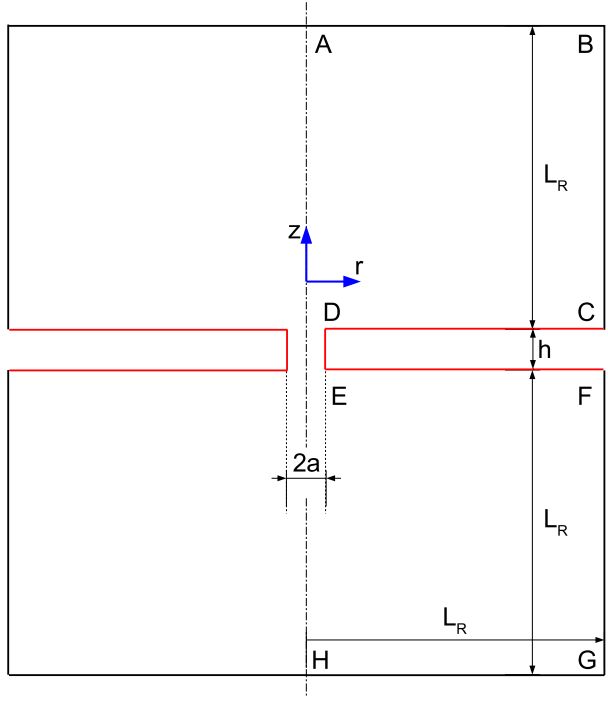}
\caption{A sketch of the axisymmetric geometry used in the simulation.}
\label{fig:system}
\end{figure}
\begin{figure}
\centering
\includegraphics[width = 1.0\textwidth]{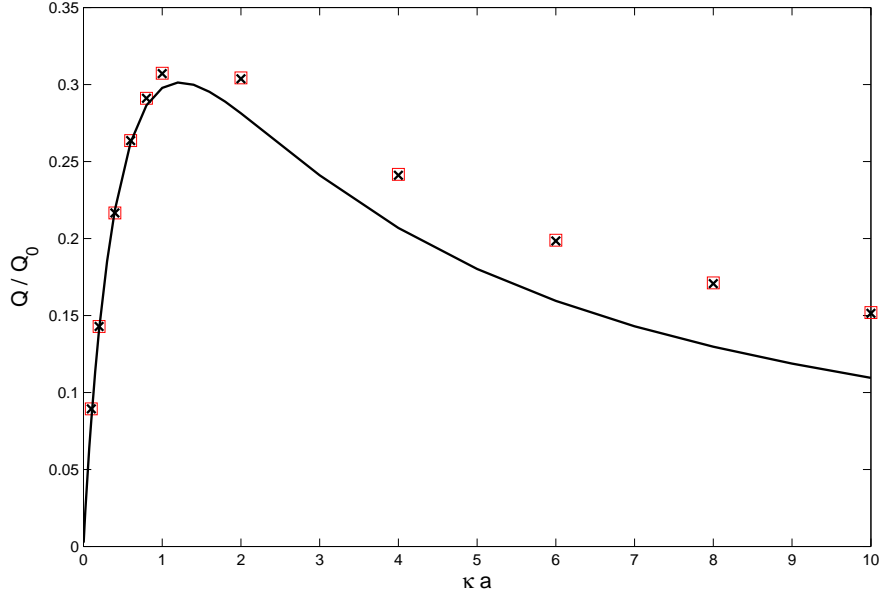}
\caption{The normalized flow rate through a circular pore of radius $a$ in a thin membrane 
as a function of $\kappa a$. The solid line shows results for a membrane of zero thickness,
obtained via the reciprocal theorem and (\ref{eq:Q2}).
The symbols are from the full numerical simulation with a membrane
of thickness $h = 0.1a$.
Squares show results for
 a non-polarizable membrane $\epsilon_s=0$, crosses show results for 
 a membrane with a dielectric constant $3.9$. The  dielectric constant of 
 the electrolyte is $80$. The effect on the flow rate
due to membrane polarizability and consequent ICEO
 is seen to be negligible.}
\label{fig:QaKappaEpw0}
\end{figure}

\section{Numerical solution of the PNP-Stokes equations}

\subsection{Numerical scheme}
 An electrohydrodynamic solver was developed to solve the Poisson--Nernst--Planck--Stokes (PNP--Stokes) system of equations using the finite volume method. The solver was based on the OpenFOAM CFD library \citep{OPENFOAM}, a C++ library designed for computational mechanics,
containing a collection of object--oriented classes developed to represent mesh, fields, matrices and the necessary operations on fields and tensors. It also provides functions to handle finite volume discretization and matrix equation solving. 

The time--independent PNP--Stokes equations are:
\begin{eqnarray}
\epsilon \nabla^2 \phi + \sum_{i=1}^{N} z_ien^i & = & 0,
\label{eq:poisson}
\\
\nabla\cdot\left\lbrack n^i\mathbf{u} -\omega^i(kT\bnabla
n^i + ez_in^i\bnabla\phi) \right\rbrack&=&0 ,
\label{eq:NP}
\\ 
-\nabla p + \mu \nabla^2 \mathbf{u} -  \nabla \phi \sum_{i=1}^{N} z_ien^i & = & 0, \label{eq:stokes}\\
\nabla \cdot \mathbf{u} & = & 0. \label{eq:continuity}
\end{eqnarray} 

In our simulation we consider a 1--1 symmetric electrolyte solution containing ions with equal mobilities. The boundary conditions to be satisfied by the solution are discussed in \S~\ref{sec:math_model}.

We apply the following scheme to solve the PNP--Stokes equations. We start from a zero flow field. Equations  (\ref{eq:poisson}) and (\ref{eq:NP}) are solved sequentially in a loop with under-relaxation until the absolute residual is smaller than $10^{-6}$. Under--relaxation is necessary because the PNP system is non--linear. The electric volume force $- \nabla \phi\sum_i z_ien^i $ is obtained from this solution and used explicitly in the next step: the solution of the incompressible Stokes flow: (\ref{eq:stokes}) and (\ref{eq:continuity}). The SIMPLE algorithm is used with a fixed volume force density. The flow field is then substituted into (\ref{eq:NP}). The PNP equations are then solved again using the updated flow field. An outer loop is constructed to iterate over the PNP loop and Stokes flow module.

For the finite volume discretization of the governing equations, central differences are used for all diffusive terms in (\ref{eq:NP}) and viscous terms in (\ref{eq:stokes}). A second--order upwind scheme is used for the convective terms in (\ref{eq:NP}). The discretized linear system is solved using a pre-conditioned conjugate gradient solver if the matrix is symmetric or a pre-conditioned bi--conjugate gradient solver if the matrix is asymmetric. The details of the numerical algorithm are given by \citet{ferziger&peric}. 

\subsection{\label{sec:math_model}Mathematical model of the nanopore}

A schematic view of the axisymmetric geometry used for the full numerical
simulations is provided in Figure~\ref{fig:system}. It consists of a circular hole of radius $a$ in a solid dielectric membrane CDEF of arbitrary thickness $h\ge 0$. The membrane surfaces CD, DE and EF have a uniform surface charge density $\sigma$. Two large cylindrical reservoirs are connected to the pore, one at each end. The length and radius of both the reservoirs are identical, and are
$L_R=\max(10a, 10\kappa^{-1})$, chosen to be much larger than either the hole radius $a$ or the Debye length $\kappa^{-1}$  in order to approximate an infinite reservoir.

We adopt the following boundary conditions \citep{Mao2013}. The ion number densities on AB and GH are constant, and equal to the number density $n_\infty$ in the bulk solution far from any charged surfaces. The electrical potentials are uniform on AB and on GH, with a potential difference of $\Delta\phi$ between the top (AB) and bottom (GH). The pressure $p_{\infty}$ on AB is uniform and equal to that on GH. On the side walls BC and FG, the radial electric field, ionic flux and radial velocity, which decay away from the pore, are set to zero. A zero tangential shear stress is imposed on flow parallel to the side walls. At the membrane surfaces CD, DE and EF, a no--flux condition is used for 
(\ref{eq:NP}), a no--slip condition for the flow; the electric field $\mathbf{E}$ undergoes a jump across the solid-fluid interface such that
$\epsilon \mathbf{E} \cdot  \hat{\mathbf{n}} - \epsilon_{s} \mathbf{E}_{s} \cdot \hat{\mathbf{n}} = \sigma$ 
where $\epsilon$ is the electrical permittivity of the
fluid and $\epsilon_{s}$ is the permittivity of the membrane, $\mathbf{E}_{s}$ is the electric field 
at the interface within the membrane and $\hat{\mathbf{n}}$ is the unit normal at the surface directed 
into the fluid. The potential is continuous across the interface. 

The strength of the applied field and the amount of surface charge can be characterized 
by the dimensionless parameters $\hat{\phi} = |\epsilon \Delta \phi/(\sigma a)|$ and 
$\hat{\sigma} = a e |\sigma | / (\epsilon kT)$ respectively. In the simulations presented here, 
the values of these parameters were kept fixed at $\hat{\phi}=0.071$ and $\hat{\sigma} = 0.273$.
The flow rate $Q$ was obtained by numerically integrating the $z$--component of the velocity over the plane $z=0$.

\subsection{Effect of membrane polarizability}
If the membrane polarizability is sufficiently small that
$|\epsilon_{s} \mathbf{E}_{s} \cdot \hat{\mathbf{n}}|\ll|\epsilon \mathbf{E} \cdot  \hat{\mathbf{n}}|$, then the jump condition of the normal component of the field may be replaced by
 $\epsilon \mathbf{E} \cdot \hat{\mathbf{n}} = \sigma$. 
In this case, the computational domain may be restricted to include only the fluid phase. This approximation was adopted for the results presented in Figures~\ref{fig:QaKappa} and
\ref{fig:flowConduc}.
Thus, effects due to Induced Charge Electroosmosis (ICEO) were neglected. 
The results of a calculation to test the validity of this assumption in the parameter range 
of interest are shown in Figure~\ref{fig:QaKappaEpw0}. The data from Figure~\ref{fig:QaKappa} for a non-polarizable membrane of thickness $h = 0.1a$ are reproduced in Figure~\ref{fig:QaKappaEpw0}. 
 For comparison, the result of a second calculation in which the dielectric constant of 
the membrane material was set to $3.9$ (corresponding to silica) is also shown. The electrolyte 
is considered polarizable with a dielectric constant of $80$. It is seen that 
the effect of membrane polarizability on the flow rate is negligible.
\bibliographystyle{jfm}

\end{document}